\documentclass[aps,prl,onecolumn,preprint,superscriptaddress,english]{revtex4-1}
\usepackage[utf8]{inputenc}
\usepackage[T1]{fontenc}
\usepackage[english]{babel}
\usepackage{bm}
\usepackage{graphicx}
\usepackage{float}
\usepackage{wrapfig}
\usepackage[dvipsnames]{xcolor}
\usepackage{placeins}
\usepackage[linktocpage=true,colorlinks=true,pdfborder={0 0 0},linkcolor=blue,citecolor=blue,filecolor=yellow,urlcolor=blue,bookmarks,pdfauthor={},]{hyperref}
\usepackage{amsmath}
\newcommand{\UCSD}{Department of Nanoengineering, University of California San Diego, La Jolla, CA 92037, USA}
\newcommand{\LasVegas}{Department of Physics \&{} Astronomy, University of Nevada Las Vegas, Las Vegas, Nevada 89154, USA}
\newcommand{\Argonne}{X-ray Sciences Division, Argonne National Laboratory, Lemont, Illinois, 60439, USA}

\babelprovide{english}

\begin{document}

\author{Daniel~Schacher}
 \affiliation{\LasVegas}
\author{Tod~A.~Pascal}
 \email{tpascal@ucsd.edu}
 \affiliation{\UCSD}
\author{Keith~V.~Lawler}
 \email{keith.lawler@unlv.edu}
 \affiliation{\LasVegas}
\author{Craig~P.~Schwartz}
 \email{craig.schwartz@unlv.edu}
 \affiliation{\LasVegas}
 \affiliation{\Argonne}

\date{\today{}}

\title{The Surface Sensitivity of X-ray Second Harmonic Generation as a Function of Energy}

\begin{abstract}
The surface sensitivity and probe depth in the x-ray regime of diamond for second harmonic generation (SHG) was investigated both analytically and computationally with velocity gauge real-time time-dependent density functional theory (VG-RT-TDDFT), which includes a full multipole expansion.
This was accomplished using two different approaches, by changing the number and location of layers that can generate SHG computationally and by exploiting the symmetry of a crystal, a similar pattern emerged. 
We find that by 1000\,eV, well above the ~285\,eV of the C $K$-edge, the SHG of diamond is dominated by the bulk, quadrupole response, in agreement with our analytic calculations.
The bulk response continues to grow as the energy is increased, becoming overwhelming by 7000\,eV.  
Near the C $K$-edge the measurement is quite surface sensitive, however, this surface sensitivity reduces as the energy increases such that by 1000\,eV (and certainly by 3500\,eV) SHG is largely bulk sensitive. 
Moreover, we find that the specific details of the crystal orientation (i.e., comparing a (001)-terminated and (111)-terminated surface) appear to have significant effects on the surface sensitivity.
\end{abstract}

\maketitle

\section{Introduction}

Following the invention of the laser~\cite{Townes}, one of its earliest and most revealing impacts was to drive optical response into regimes where symmetry considerations dominate.  
The large electromagnetic fields enabled the first demonstrations of second harmonic generation (SHG) in bulk noncentrosymmetric crystals~\cite{Franken}, and such nonlinear effects soon became foundational tools in laser science, optics, and spectroscopy~\cite{Boyd}.  
In second harmonic generation (SHG), two photons of equal energy cooperate to yield one of doubled energy; in the closely related sum-frequency generation (SFG), two inequivalent photons combine to produce a third whose energy is the sum of the two inequivalent photons.  
In the electric-dipole limit, these processes are exquisitely sensitive to symmetry: they generate signal only where inversion symmetry is broken, requiring both microscopic and macroscopic departures from centrosymmetry~\cite{Boyd,Kleinman,Petersen}.
Early reports of SHG from nominally centrosymmetric bulk media were therefore puzzling; the resolution came from recognizing that bulk dipole responses vanish by symmetry and that the observed ``bulk'' signals arose under conditions that activated higher-order multipole channels, a distinction clarified in subsequent analyses~\cite{Bloembergen,Guyot}.  

This clarification highlights a practical constraint: the surface specificity of SHG in centrosymmetric media rests on the assumption that higher-order terms---most notably the electric-quadrupole contribution---remain small.
The electric-quadrupole term, however, can be present in the bulk, and over a finite propagation length its contribution can accumulate and overwhelm the comparatively weak surface signal.
Indeed, even at visible and infrared wavelengths, bulk SHG can be generated via specific higher-order transitions~\cite{Akihiro,Akihiro2,taka-aki}.  
This is not a minor complication for interfacial spectroscopy: to interpret SHG as a probe of surface structure, the surface contribution must remain appreciable.  
The challenge becomes more acute at shorter wavelengths, where higher photon energies strengthen bulk multipole pathways.  

For a long period, this short-wavelength problem drew little attention because nonlinear optics was largely confined to the visible and infrared spectral regions~\cite{Boyd}, a practical limitation set by the photon energy of sufficiently intense light sources~\cite{keller_recent_2003,keller_ultrafast_2021}.  
That landscape changed with the development of x-ray free-electron lasers (XFELs)~\cite{Bostedt}.  
These machines deliver intense, ultrafast x-ray pulses that open access to nonlinear interactions on x-ray energy scales~\cite{Marinelli}.  
As a result, multiple nonlinear x-ray processes---including four-wave mixing and two-photon absorption---have now been demonstrated~\cite{Reistpa,bencivenga_four-wave_2015}, along with x-ray and optical mixing (XOM)\cite{Chance}, hard x-ray SHG and SFG, and soft x-ray SHG~\cite{Lam2018PRL,Shwartz2014PRL,Glover2012Nature,Hutchison2024JPCB}.  
Owing to their broad tunability and unprecedented field strengths, XFELs make feasible measurements that were previously out of reach.  

Hard x-ray SHG departs fundamentally from its visible and infrared counterparts.  
Long before direct observation, Freund and Levine argued that at \textup{\AA}ngstr{\"o}m wavelengths, well above core resonances, the dominant nonlinearity should resemble a plasma-like response of nearly free electrons~\cite{Freund1969PRL}. 
This was subsequently supported by the observation of x-ray parametric down-conversion at hard x-ray energies~\cite{Eisenberger1971PRL}.  
Only with the advent of XFELs could these properties of hard x-ray SHG be confirmed experimentally~\cite{Glover2012Nature,Shwartz2014PRL}.  
In this regime, where photon energies greatly exceed core binding energies, the driven electrons behave quasi-freely, and the nonlinear response is governed primarily by spatial inhomogeneities of the electron density, ${\rho} (\mathbf{r})$~\cite{Shwartz2014PRL,Freund1969PRL}.  
Because ${\rho}(\mathbf{r})$ varies on length scales comparable to the photon wavelength at these energies, the SHG signal becomes increasingly bulk-sensitive to structural parameters---interplanar spacings, bond lengths---while the surface-induced dipole contribution recedes with rising photon energy.  
In centrosymmetric media, nonuniform electron density and ponderomotive forces then act as bulk sources of the second-harmonic current~\cite{Freund1969PRL}.  
Despite its bulk character, hard x-ray SHG requires intense fields: in diamond, it has been observed at a $7.3\,\mathrm{keV}$ pump, with measured conversion efficiencies of $5.8\times10^{-11}$ under intensities on the order of $10^{16}\,\mathrm{W\,m^{-2}}$, and the emitted signal scales quadratically with the fundamental pulses' energy under phase-matched conditions~\cite{Shwartz2014PRL,Yudovich2015JOSAB}.  

The soft x-ray region provides a revealing testbed for the crossover from surface to bulk sensitivity in centrosymmetric media.  
Near core-level resonances, recent reports suggest that SHG recovers sensitivity to inversion-symmetry breaking---acquiring pronounced interfacial character while retaining the element specificity typical of x-ray probes~\cite{Lam2018PRL,Berger2021EUVSHG,stohr_nexafs_2003}.  
A faithful theoretical description in this intermediate regime necessarily expands the total polarization beyond the dipole term, with the second-harmonic response comprising surface/interface-induced electric dipoles together with bulk electric-quadrupole and magnetic-dipole contributions~\cite{Shen1989Nature}.  
Previous combined theory--experiment near the carbon $K$-edge showed that the SHG response in graphite thin films decays approximately exponentially with depth: first-principles calculations (supported by experiment) attribute about $63\%$ of the net $\chi^{(2)}$ to the top atomic layer, and more than $95\%$ to the top three layers~\cite{Lam2018PRL}.  
However, such analyses often rely on density-functional perturbation theory to compute an effective $\chi^{(2)}$ response~\cite{Lam2018PRL,Berger2021EUVSHG}, an approach that adopts the dipole approximation and thus neglects higher-order field terms that generate bulk signal (e.g., electric-quadrupole and magnetic-dipole).  
A more complete determination---one that properly accounts for multipole effects---is therefore required to identify, in energy, where SHG transitions from surface-dominated to bulk-dominated behavior.  

While higher-order multipoles are frequently neglected at longer wavelengths (IR, visible), that simplification has not always been justified~\cite{Akihiro2}.  
It becomes increasingly untenable as wavelengths shorten, because electric-quadrupole and magnetic-dipole pathways are bulk-allowed in centrosymmetric media---certainly by hard x-ray energies.  
Quantifying their contributions in the intermediate (soft x-ray) regime is thus essential, as the crossover between resonant, symmetry-sensitive interface mechanisms and plasma-like bulk mechanisms remains incompletely mapped.  
Here, by combining computational and analytical approaches, we chart this transition for diamond, with particular attention to the onset and scaling of bulk contributions to SHG as functions of x-ray wavelength and material parameters.  

%\section{Formalism}

Following the perturbative treatment of \citet{Freund1969PRL} using the notation of \citet{KubotaTamsaku2023NonlinearXRay}, the interaction of an electron with an electromagnetic field can be decomposed into contributions associated with the $\mathbf{p}\cdot\mathbf{A}$ and $\mathbf{A}^{2}$ terms in the Hamiltonian where $\mathbf{A}$ represents the vector potential and $\mathbf{p}$ the momentum.
Here our unperturbed Hamiltonian $\mathcal{H}_0$ is:
\begin{equation}
\mathcal{H}_0 = \sum_i \left[ \frac{\mathbf{p}_i^{\,2}}{2m} + V(\mathbf{r}_i) \right],
\end{equation}
where $V(\mathbf{r}_i)$ is the external potential and $\mathbf{p}_i$ is the momentum for the $i$th electron, and $m$ is the mass of the electron. Our perturbation Hamiltonian $\mathcal{H}'$ containing the vector potential terms is:
\begin{equation}
\mathcal{H}' = \sum_i \left[ \frac{e}{mc}\,\mathbf{p}_i \cdot \mathbf{A}(\mathbf{r}_i,t)
+ \frac{e^{2}}{2mc^{2}}\,\mathbf{A}^{2}(\mathbf{r}_i,t) \right],
\end{equation}
where $e$ is the charge of the electron, $c$ is the speed of light in a vacuum, and $\mathbf{A}(\mathbf{r}_i,t)$ refers to the vector potential at position $\mathbf{r}_i$ and time $t$ for the $i$th electron.
When we consider the two x-ray fields associated with a second-order nonlinear signal, the perturbation Hamiltonian becomes:
\begin{equation}
\mathcal{H}' = \frac{e}{mc}\,\mathbf{p} \cdot
\left\{\mathbf{A}_{1}(\mathbf{r},t) + \mathbf{A}_{2}(\mathbf{r},t)\right\}
+ \frac{e^{2}}{2mc^{2}}
\left\{\mathbf{A}_{1}(\mathbf{r},t) + \mathbf{A}_{2}(\mathbf{r},t)\right\}^{2}.
\end{equation}
Then via standard perturbation theory, when considering two x-ray fields with wave vectors and angular frequency with $\mathbf{K}_{1},\omega_{1}$ and  $\mathbf{K}_{2},\omega_{2}$ respectively, we arrive at the non-linear current density $\mathbf{J}_{3}$ with associated wave vector and angular frequency with $\mathbf{K}_{3},\omega_{3}$:
\begin{equation}
\mathbf{J}_{3}(\mathbf{K}_{3},\omega_{3})
 = - i \pi \omega_{3}\,
 \tilde{\boldsymbol{\beta}}(\mathbf{K}_{3}-\mathbf{K}_{1}-\mathbf{K}_{2},
   \omega_{1},\omega_{2})
 : \boldsymbol{\epsilon}_{1}\boldsymbol{\epsilon}_{2}\,
   E_{1}E_{2}\,
   \delta(\omega_{3}-\omega_{1}-\omega_{2}),
\end{equation}
where $\tilde{\boldsymbol{\beta}}$ is the Fourier transform of the second-order nonlinear polarizability ($\chi^{(2)}$), and $\boldsymbol{\epsilon}_{1}$ and $E_{1}$ and $\boldsymbol{\epsilon}_{2}$ and $E_{2}$ are the polarization vectors and field strengths associated with their respective x-ray fields.
Note that for SHG, the two incoming fields will be equivalent, i.e. $\mathbf{K}_{1},\omega_{1},\boldsymbol{\epsilon}_{1},E_{1}$ = $\mathbf{K}_{2},\omega_{2},\boldsymbol{\epsilon}_{2},E_{2}$.
From this analysis, one obtains a frequency-dependent second-order polarizability
\begin{equation}
\tilde{\boldsymbol{\beta}}(\mathbf{K}_{3}-\mathbf{K}_{1}-\mathbf{K}_{2},
  \omega_{1},\omega_{2})
 : \boldsymbol{\epsilon}_{1}\boldsymbol{\epsilon}_{2}
 = \frac{i e^{3}}{m^{3}\hbar^{2}\omega_{1}\omega_{2}\omega_{3}}\,
 (\mathcal{U} + \mathcal{B}),
\end{equation}
where $\mathcal{B}$ arises solely from $\mathbf{p}\cdot\mathbf{A}$ perturbations 
and $\mathcal{U}$ from perturbations including $\mathbf{A}^{2}$;  ${\hbar}$ is Planck's constant divided by $2\pi$~\cite{Freund1969PRL,Shwartz2014PRL}.
In the long-wavelength regime (optical), $\mathcal{B}$ dominates, whereas at hard x-ray energies $\mathcal{U}$ becomes the leading contribution as the charge density contribution in the $\mathbf{A}^{2}$ terms are expected to be the dominant off-resonant signals \cite{doi:10.1021/acs.jctc.9b00346}. 
Since $\mathcal{B}$ vanishes in centrosymmetric media within the electric dipole approximation, it is only nonzero when inversion symmetry is broken, while $\mathcal{U}$ is always allowed and therefore provides a bulk contribution even in centrosymmetric crystals~\cite{Shen1989Nature,Shwartz2014PRL}. Thus, the crossover between surface and bulk sensitivity in a centrosymmetric medium will be governed by the relative magnitudes of the $\mathcal{B}$ and $\mathcal{U}$ terms.

A different approach to quantify the surface sensitivity was undertaken by Mizrahi and Sipe~\cite{Mizrahi1988SurfaceSHG} beginning from longer wavelength approximations. They also emphasize that surface and bulk contributions interfere in the detected second harmonic (SH) signal and that the effective bulk contribution is depth-limited by the phase-mismatch (“coupling depth”). As a function of energy, there may be nontrivial phase evolution and partial saturation with depth, leading to a featured spectrum rather than one that changes monotonically.
In that work, they derive an expression for the coupling depth of the material that effectively contributes to the bulk SHG.
This coupling depth ($\xi$) with grazing incident angles is proportional to the fundamental pulses' wavelength ($\lambda$):
\begin{equation}
\xi \;\propto\; \frac{\lambda}{8}\,
\end{equation}

There are also symmetry relations which can be used, and this has been worked out in the SHG of silicon surfaces, which has an identical symmetry to diamond~\cite{Tom}. For a (001)-terminated surface with $4m$ symmetry, there are only three surface allowed components of the $\chi^{(2)}$ tensor which are $\chi^{(2)}_{ZZZ}$, $\chi^{(2)}_{ZXX}$, and $\chi^{(2)}_{XZX}$ (here $Z$ is perpendicular to the surface, and $X$ and $Y$ are arbitrary and interchangeable). This is utilized here as when other $\chi^{(2)}$ tensor components are found to be non-zero, they can be automatically assigned to bulk phenomena. 

%\section{Methods}

\begin{figure}[h]
    \centering
    \includegraphics[width=80mm]{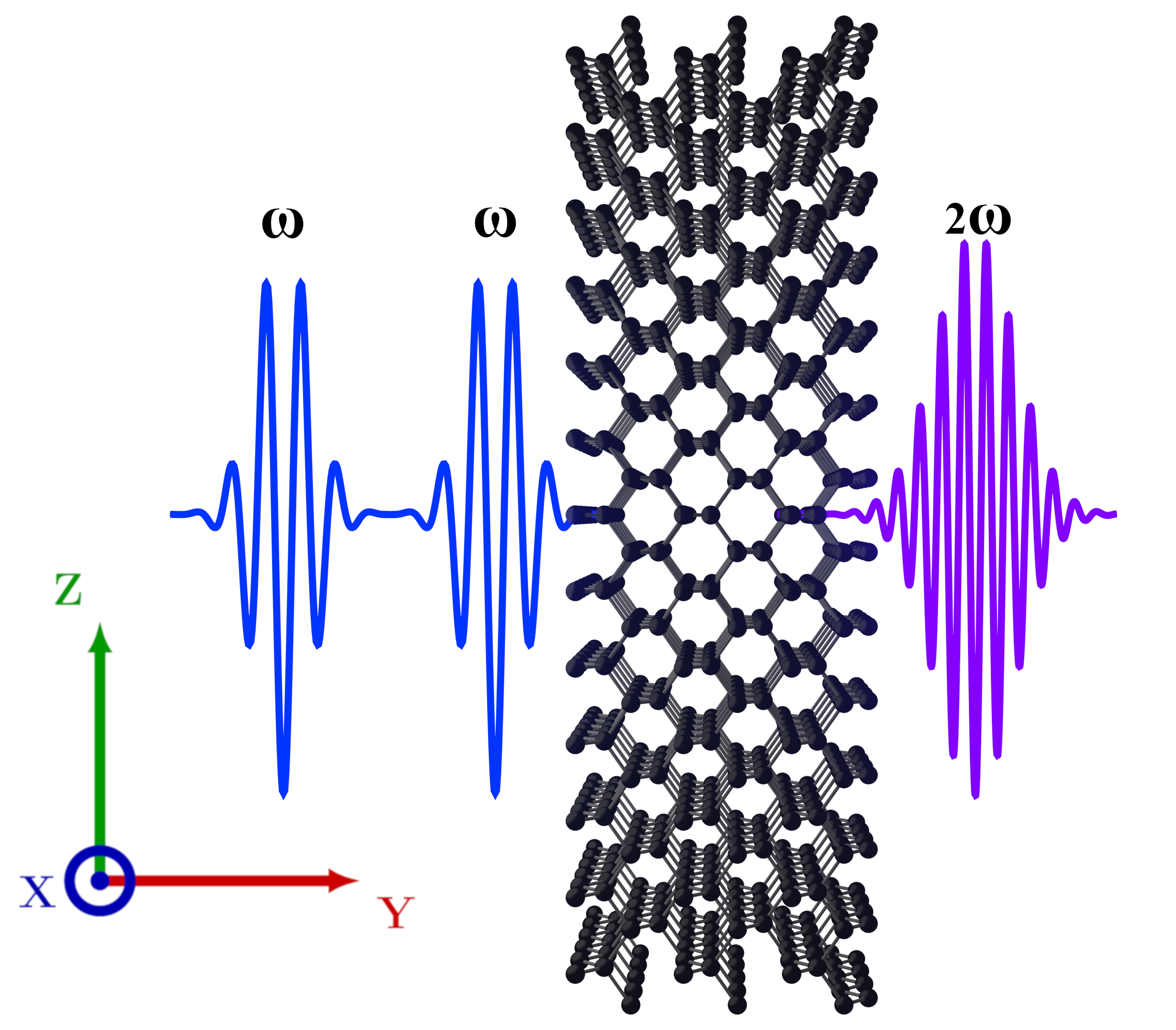}
    \caption{A conceptual drawing of what is being calculated.  An electric field with polarization shown in Z is incident on a (001) or (111)-terminated slab of diamond. The current density in the diamond is then calculated and used to monitor SHG.} 
    \label{fig:cartoon}
\end{figure}

The surface sensitivity is also tested here with first-principles simulations that are performed using the velocity-gauge real-time time-dependent density functional theory (VG-RT-TDDFT) formalism implemented in a customized version of the \textsc{SIESTA} code~\cite{Soler2002JPCM,Pemmaraju2018CPC}. 
VG-RT-TDDFT has been shown to provide accurate predictions for soft x-ray SHG spectra. \cite{hoffman_2025,woodahl_probing_2023,helk_table-top_nodate}
In this methodology, the laser field is represented by a time-dependent vector potential. 
The response is evaluated from the current density of the system, so that all contributions from the multipole expansion are included, in particular bulk responses arising from electric-quadrupole and higher order moments.
The calculations employ a double-$\zeta$ plus polarization basis of localized numerical atomic orbitals. A norm-conserving Troullier--Martins pseudopotential explicitly pseudizing up to the $\mathrm{C}:\{1s,2p,3d\}$ states and one pseudizing up to the $\mathrm{C}:\{2s,2p,3d\}$ states were used to appropriately model the core states and relevant transitions~\cite{TroullierMartins1991PRB,Hamann2013PRB}.
By controlling whether the pseudopotential for a given atom includes the $1s$ state in the valence, we can effectively switch on or off its core-level response, which allows us to isolate the contribution from specific layers/atoms in the slab. 
This is important as the SHG contribution from the bottom surface will exactly cancel that from the top surface in our finite system size, so the $1s$ states in the bottom half of the slab are always treated as in the pseudized core in our simulations. 
Within the top half of the slab, the particular layers with $1s$ states in the valence are varied to calculate depth dependent SH responses. 
A cartoon showing what is being calculated is shown in Fig.~\ref{fig:cartoon}. 
This artificial symmetry breaking in the middle of the slab is expected to have minimal effects on the valence and conduction states of the system affecting the SHG response as the C $1s$ states can be treated as atom-localized and any asymmetric polarization effects depend on how closely the pseudized and all-electron C $1s$ states match.
The pseudized and all-electron states potentials are designed here to have as close an electronic structure as possible.

Exchange--correlation effects are described within the generalized-gradient approximation (GGA) using the PBE functional~\cite{Perdew1997PBE}.
We modeled diamond (C, mp-66, cubic lattice constant of 3.567~$\text{\AA}$) \cite{MPmp66} as both (100) and (111)-terminated slabs, using $1\times 1\times 8$ unit cells separated by a $10~\text{\AA}$ vacuum region (called here $1\times 1\times 16$ layers).
A layer was defined as half of the carbon atoms in one of the unit cells used such that there are two layers per unit cell.
For the (001)-terminated slabs, a tetragonal unit cell with 1 atom per layer was used; for the (111)-terminated slabs, a hexagonal unit cell with 2 atoms per layer was used; and for the bulk simulations, a conventional cubic unit cell with 8 total atoms was used.
The Brillouin zone was sampled with a $10\times 10\times 1$ Monkhorst-Pack grid \cite{MonkhorstPack1976PRB}; bulk calculations used equivalent $k$-point densities in the three-dimensional Brillouin zone.

The velocity gauge was used to apply external fields via a spatially uniform vector potential $\mathbf{A}(t)$ with polarization perpendicular to the surface, taken either as a $\delta$-kick for broad spectrum linear signal calculations or as a finite-duration sine pulse with a sine$^{2}$ envelope for nonlinear signal calculations.
We considered peak intensities spanning $10^{15}$--$10^{22}~\mathrm{W\,cm^{-2}}$ and photon energies from 282\,eV  to 7\,eV.
The time step $\Delta t$ was chosen to satisfy the Nyquist criterion for a target spectral bandwidth up to photon energy $E_{\max}$,
\begin{equation}
  E_{\max} = \hbar \omega_{\max}
  = \frac{\hbar \pi}{\Delta t}
  \quad\Rightarrow\quad
  \Delta t \le \frac{\hbar \pi}{E_{\max}}.
\end{equation}
For $E_{\max}=15$\,keV this bound gives
$\Delta t \lesssim 1.4\times 10^{-19}~\mathrm{s}$. 
Consequently, we used attosecond-scale steps of approximately $6.65\times 10^{-5}~\mathrm{fs}$ and a total propagation time $T = 1~\mathrm{fs}$. 
For selected runs, a $\delta$-kick perturbation was used to obtain linear absorption spectra, following standard real-time TDDFT practice \cite{Yabana1996PRB}.

The macroscopic current density $J(t)$ was computed from the Kohn--Sham states in the velocity gauge and Fourier transformed to the frequency domain to obtain $J(\omega)$. For SHG, it is convenient to use the formalism $\omega_{1}=\omega_{2}=\omega_{0}$ exploiting both input photons having the same energy and calling that $\omega_{0}$.  This naturally gives $\omega_{3}=\omega_{1}+\omega_{2}=2\omega_{0}$ where $\omega_{3}$ is the signal.
To extract the second-harmonic response, we evaluated the current at the second-harmonic frequency $2\omega_{0}$ and fit it to the second-order quadratic relation
\begin{equation}
  J_{\beta}(2\omega_{0}) =
  \chi_{\beta\alpha\alpha}^{(2)}(2\omega_{0};\omega_{0},\omega_{0})\,E_{\alpha}^{2}(\omega_{0}),
  \label{eq:secondOrder}
\end{equation}
where $E(\omega_{0})$ is the Fourier amplitude of the driving field at the fundamental (incident) photons' frequency (i.e. the input field strength), and $\alpha$ and $\beta$ are the polarization directions of the fundamental pulse and the response respectively. 
The $J(2\omega_{0})$ values were then plotted versus the fundamental's field strength and fitted to Eqn.~\ref{eq:secondOrder}; deviations from quadratic scaling at the highest intensities were discarded to avoid the influence of other phenomena. As expected, the resulting curvers are well fit by the second-order relation and thus is used to obtain $\chi^{(2)}$. This process was repeated by varying intensities ($10^{12}$ to $10^{22}~\mathrm{W\,cm^{-2}}$) and  photon energies.  A minimum of 3 different nonzero intensities were used per energy.  At higher intensities, the second-harmonic signal eventually deviates from quadratic behavior and increases sharply, consistent with the onset of saturation and ionization processes~\cite{Shwartz2014PRL,Yudovich2015JOSAB}. To avoid these artifacts, we restrict our analysis to the intensity range over which the response remains well described by a quadratic fit and where the extracted $\chi^{(2)}$ is maximized for a given photon energy. The current density at the fundamental frequency, $J(\omega_{0})$ was also plotted against the fundamental pulses' field strength to verify that it follows the appropriate linear relationship $J_{\beta}(\omega_{0}) = \chi_{\beta\alpha}^{(1)}(\omega_{0};\omega_{0}) E_{\alpha}(\omega_{0})$.

%\section{Results}

\begin{figure}[h]
    \centering
    \includegraphics[width=0.45\columnwidth]{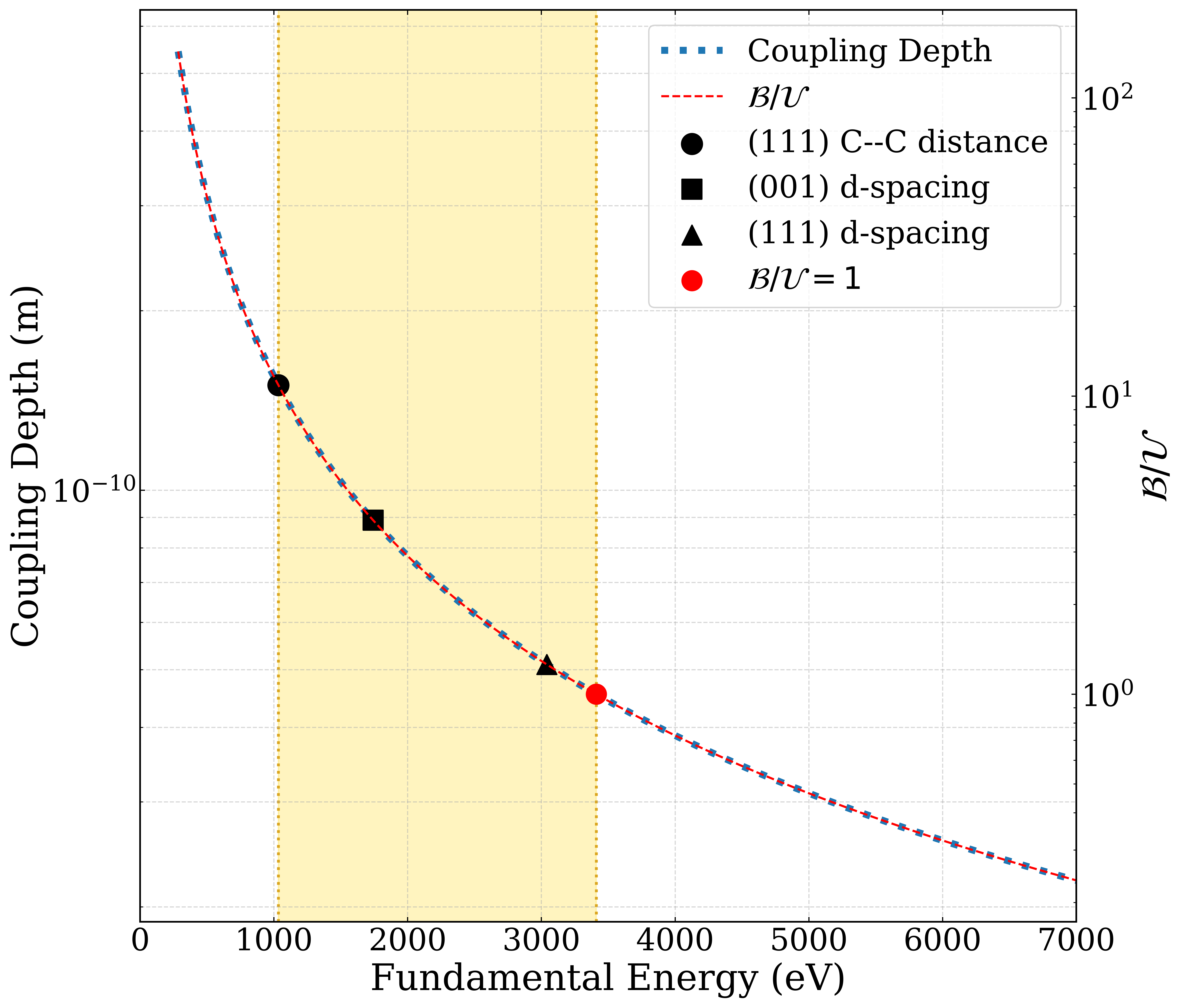}
    \caption{Plot of coupling depth and |$\mathcal{B}$/$\mathcal{U}$| as a function of energy. The left axis is the effective coupling depth according to the formalism of Mizrahi and Sipe~\cite{Mizrahi1988SurfaceSHG}. The right axis is the value of the ratio of |$\mathcal{B}$/$\mathcal{U}$|~\cite{Freund1969PRL}. Points of interest are marked where black points represent wavelengths corresponding to important distances in diamond and the red circle corresponds to |$\mathcal{B}$/$\mathcal{U}$| being unity. The marked points use the coherence length with distances associated with the (111) layer separation, the (001) layer separation and the distance between two adjacent carbon atoms denoted with a black circle, triangle and square respectively. The yellow region corresponds to where one therefore might expect bulk contributions to become dominant.} 
    \label{fig:matsuda_ratio}
\end{figure}

In order to determine the interface sensitivity, we begin with the following derivation laid out by \citet{KubotaTamsaku2023NonlinearXRay}. They arrived at approximations for $\mathcal{U}$ and the first term in the perturbation expression for $\mathcal{B}$ in the x-ray regime with the assumptions that $\langle \mathbf{K} \cdot \mathbf{r} \rangle \ll 1$ and that the size of the dipoles is the Bohr radius. We use these expressions for the magnitudes of $\mathcal{B}$ and $\mathcal{U}$ to calculate a ratio that will act as a measure of the relative bulk response due to plasma nonlinearities in the x-ray region:
\begin{equation}
|\mathcal{B}|
  \sim \frac{\hbar^{6}\Omega_{\mathrm{res}}^{3}}{e^{6}\omega_{x}^{2}} ,
\end{equation}
\begin{equation}
|\mathcal{U}|
  \sim \frac{m\hbar^{2}\tilde{\rho}}{c} ,
\end{equation}
\begin{equation}
\left|\frac{\mathcal{B}}{\mathcal{U}}\right|
  \sim \frac{1}{\alpha^{3}}
       \frac{(\hbar\Omega_{\mathrm{res}})^{3}}{m c^{2} (\hbar\omega_{x})^{2}}
       \frac{1}{\tilde{\rho}}.
\end{equation}
Here $\omega_{2}$ and $\omega_{3}$ have been set to $\omega_{x}$ for x-ray frequencies, $\Omega_{\mathrm{res}}$ is the major resonant frequency, $\tilde{\rho}$ is the Fourier transform of the electron density, $\alpha$ is the fine structure constant and the rest is as defined earlier. The properties of a specific crystal thus can have an effect electronically through $\Omega_{\mathrm{res}}$ and structurally through $\tilde{\rho}$. Using this framework, we estimate the relative magnitudes of $\mathcal{B}$ and $\mathcal{U}$ in the x-ray regime and plot their ratio as a function of photon energy in Fig.~\ref{fig:matsuda_ratio} to determine the crossover to bulk-dominated behavior, defined by $|\mathcal{B}/\mathcal{U}| < 1$. Using the value of the carbon $K$-edge for $\Omega_{\mathrm{res}}$, and approximating $\tilde{\rho}$ as 10, the value where |$\mathcal{B}$/$\mathcal{U}$| is unity is $\sim$3401.6\,eV (Fig.~\ref{fig:matsuda_ratio}, red circle), providing an upper limit for where bulk plasma-like nonlinearities are expected to become dominant. 

These results from Freund and Levine can be compared to the results for the coupling depth of Mizrahi and Sipe~\cite{Mizrahi1988SurfaceSHG, Freund1969PRL}. For the purposes of these calculations, the polarization of the electric field is perpendicular to the surface (i.e. the Z-axis) and we plot the coupling depth as a function of energy and compare it to our |$\mathcal{B}$/$\mathcal{U}$| ratio derived from the work of Freund and Levine (Fig.~\ref{fig:matsuda_ratio}). Points of specific interest are plotted---note both the ratio of |$\mathcal{B}$/$\mathcal{U}$| and the coupling depth have the same hyperbolic shape.  For the coupling depth, we plot the wavelength corresponding to the coupling depth between the (001) planes, the (111) planes, and adjacent carbon atoms in diamond. Therefore, one would expect the bulk transition to occur somewhere in or near the shaded yellow region, corresponding to the minimum and maximum values where these calculations suggest a change in behavior in diamond.  

These calculations were thus used to guide the RT-TDDFT calculations. To validate our electronic structure, a real-time trajectory with a $\delta$-kick impulse was used to compute the bulk linear absorption spectrum. The resulting spectrum, shown in the Supporting Information, exhibits good agreement with an edge onset of $\sim$285.2\,eV as compared to the experimental value of $\sim$284\,eV~\cite{Ma1992SoftXrayRIXS}.

\begin{figure}[H]
    \centering
    \includegraphics[width=1\columnwidth]{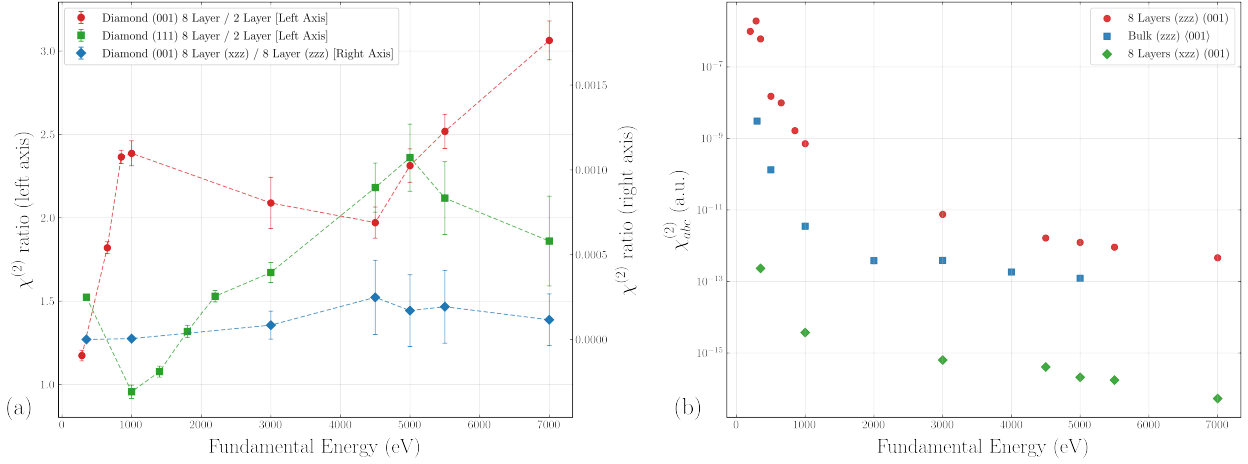}
    \caption{(a) Ratio of total $\chi^{(2)}_{ZZZ}$ for 8 active layers versus 2 layers in both (001)-terminated (red circles) and (111)-terminated (green squares) diamond slabs [left axis]. Ratio of $\chi^{(2)}_{XZZ}$ versus $\chi^{(2)}_{ZZZ}$ for 8 active layers in a (001)-terminated diamond slab (blue diamond) [right axis]. Lines connect adjacent points. (b) Calculated $\chi^{(2)}_{ZZZ}$ as a function of energy for 8 active layers in a (001)-terminated diamond slab (red circles), $\chi^{(2)}_{ZZZ}$ for a periodic bulk of diamond in its cubic cell illuminated from the $\langle$001$\rangle$ direction (blue squares), and $\chi^{(2)}_{XZZ}$ for 8 active layers in a (001)-terminated diamond slab (green diamonds). Note, $\chi^{(2)}_{XZZ}$ should be zero in the dipole limit for the (001)-terminated diamond surface with $4m$ symmetry.} 
    \label{fig:chi2ratio}
\end{figure}

To understand surface sensitivity, the value of $\chi^{(2)}$ was computed from VG-RT-TDDFT trajectories for slabs with varied quantities of active layers, i.e. layers where the C $1s$ states are fully calculated. The ratio of the $\chi^{(2)}_{ZZZ}$ for the topmost 2 surface layers as compared to the 8 total layers of the top half of the slab for both a (001) and (111)-terminated surface of diamond is shown in Fig.~\ref{fig:chi2ratio}(a). As this is total signal, a perfectly surface sensitive signal coming from just the topmost unit cell of the slab would give a ratio of 1.0 and a perfectly bulk signal would be 4.0. We begin here with discussion of the simpler (001)-terminated surface (red circles Fig.~\ref{fig:chi2ratio}(a)) which goes from quite surface sensitive to quite bulk sensitive. Error bars are calculated from the fractional uncertainties of the quadratic fit and added in quadrature. Starting at low energy near the carbon $K$-edge, the signal is surface dominant with an overwhelming amount of the signal being due to the top two layers of the surface consistent with \citet{Lam2018PRL}. As one moves up in energy, going from near the edge to 1000\,eV, the signal from the bulk rises rapidly. 
The contribution from layers 3-8 then levels off, likely due to cancellation of signal as one transitions through the various coherence lengths as shown in Fig.~\ref{fig:matsuda_ratio}. At energies above 4500\,eV, the bulk signal becomes even more dominant with the probe becoming truly bulk by 7000\,eV, with each layer almost contributing equally to the signal consistent with \citet{Shwartz2014PRL}.
To maintain surface sensitivity in this orientation, it is essential to remain close to the resonance energy; otherwise, additional effects emerge and produce a complicated set of phenomena dictating surface sensitivity from roughly 1000--5000\,eV, before the bulk response ultimately dominates.

The (111)-terminated surface exhibits a different behavior quantitatively (green squares Fig.~\ref{fig:chi2ratio}(a)), likely due to the much more complicated number of bond distances and the $3m$ symmetry of the surface allowing an additional nonzero anisotropic tensor component, $\chi_{\xi\xi\xi}^{(2)}$ where $\xi$ is the projection of (001) onto the surface~\cite{Tom}. 
In this case, the surface sensitivity near the carbon K-edge is lower than at 1000\,eV, likely due to ionization and resonant effects near the absorption edge. The response becomes progressively more bulk-dominated above 1000 eV, reaching its most bulk-sensitive value around 5000\,eV.
As the energy further increases to 7000\,eV the trend appears to reverse slightly (i.e., the signal becomes marginally more surface sensitive), but this change is within the uncertainty, and the signal would still be bulk dominated.
Given this nonmonotonic behavior found in the (111)-terminated surface, we focus on the simpler (001)-terminated surface for the analysis that follows.

In order to provide deeper insights, the bulk response of the slab system can also be assessed by calculating its $\chi^{(2)}_{XZZ}$, which within the dipole limit should be 0 for a (001)-terminated surface of diamond due to symmetry.
These calculations are shown in Fig.~\ref{fig:chi2ratio}(b) (green diamonds) as a function of the energy of the input field.
Note that these calculations are the total SHG response and thus depend on the number of active C $1s$ states in the simulation: here using 8 atoms for the 8 layer, and 8 atoms for the pure bulk.
For the calculations of a pure, periodic diamond bulk in its cubic unit cell (blue squares Fig.~\ref{fig:chi2ratio}(b)), the amount of bulk signal does decrease with increasing energy, but this is due to more total signal at lower energy as SHG intensity intrinsically decreases with increasing energy; SHG intensity goes as the inverse wavelength squared in the non-resonant limit~\cite{Boyd, He:24}. 

Because of the decreasing SHG intensity as a function of wavelength, it is worth comparing the amount of bulk signal to the surface signal at a given wavelength. 
This is done in Fig.~\ref{fig:chi2ratio}(a) (blue diamonds) which compares the ratio of $\chi^{(2)}_{XZZ}$ versus $\chi^{(2)}_{ZZZ}$ for 8 active layers in diamond (001).
The key result of that comparison is that an energy greater than 1000\,eV the symmetry disallowed components take on a value well above zero.
The ratio of the $\chi^{(2)}_{ZZZ}$ for bulk diamond to the slab's $\chi^{(2)}_{ZZZ}$ is shown in Fig.~S3 of the Supporting Information, and this comparison shows a dramatic increase in the bulk response between 1000 and 3000\,eV.
One key difference between the two approaches for this analysis is that the comparison of the bulk to the slab takes on a roughly ten-fold increase in magnitude by 3000\,eV whereas the tensor disallowed components only take on a potentially nonzero value within uncertainty at 3000\,eV and are only well above zero by 4500\,eV.
It is worth emphasizing that the transition to bulk signal is slightly different between these two approaches, due to slightly different methods and meanings of bulk in these cases.
However, within either definition, the x-ray SHG signal is dominated by bulk effects by 3000\,eV if not at lower energy but within the yellow energy window in Fig.~\ref{fig:matsuda_ratio}, and the signal is much more surface-dominant near resonance at the carbon $K$-edge. 
The plot of the ratio of $\chi^{(2)}_{XZZ}$ versus $\chi^{(2)}_{ZZZ}$ for 8 active layers in (111)-terminated diamond slab can be found in Fig.~S4 of the Supporting Information and shows limited energy dependence and a mostly negligible magnitude.

\begin{figure}[H]
    \centering
    \includegraphics[width=1\columnwidth]{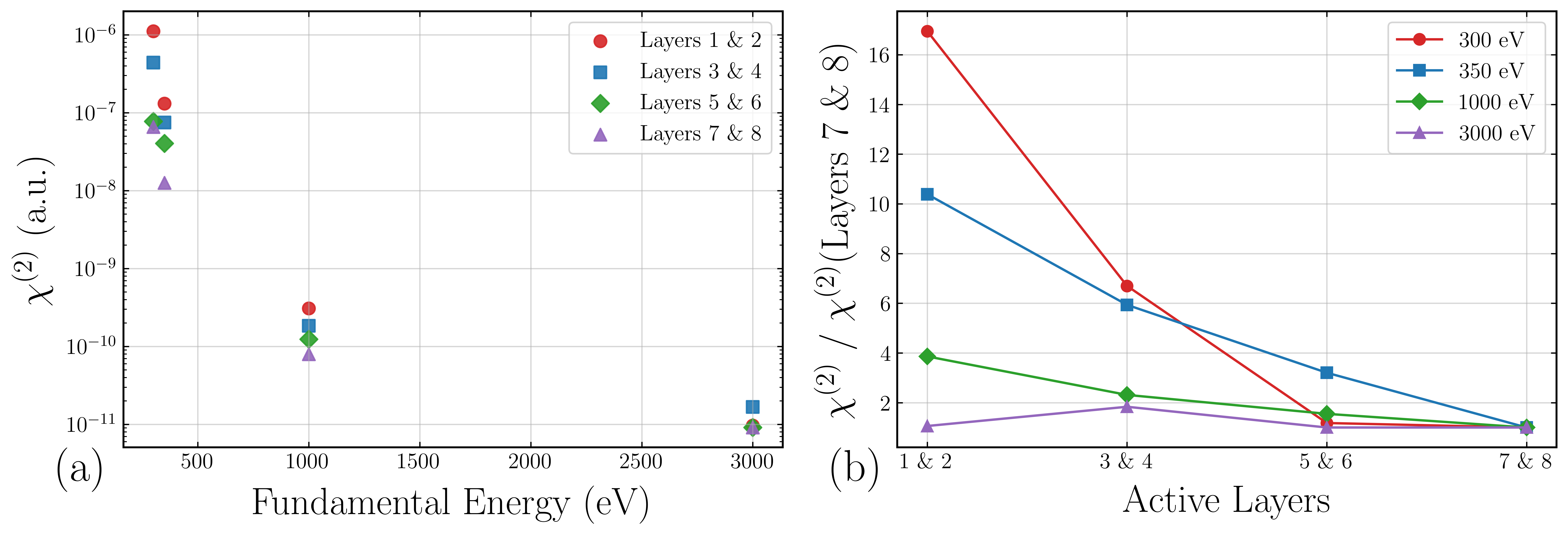}
    \caption{(a) $\chi^{(2)}_{ZZZ}$ for 2 active layers at variable depths as a function of the fundamental pulse's energy. The topmost layer is defined as layer 1. The top 2 layers (red circles), layers 3 \& 4 (blue squares), layers 5 \& 6 (green diamonds) and layers 7 \& 8 (purple triangles) are shown for energies from 300--3000\,eV.  Active layers are defined as those with the C $1s$ state in their valence.
    (b) Ratio of $\chi^{(2)}_{ZZZ}$ for active layers 1 \& 2, 3 \& 4, 5 \& 6 and 7 \& 8 normalized to the deepest active layers (7 \& 8) as a function of the incident pulses' fundamental energy. Energies shown are 300\,eV (red circles), 350\,eV (blue squares), 1000\,eV (green diamonds) and 3000\,eV (purple triangles). See text for details of these calculations and their meanings.} 
    \label{fig:buriedlayer}
\end{figure}

Another approach for quantifying the signal is undertaken in (Fig.~\ref{fig:buriedlayer}).
Here different layers were made 'active' by selecting which pseudopotential is used, the C $1s$ states are in the valence or in the pseudized core for the atoms in a specific layer.
This allows for the calculation of the contribution of specific layers to the SHG signal, while adding a potential complication from inducing a signal due to the introduction of slight symmetry breaking from the lack of active $1s$ states in certain layers --- i.e., using a pseudopotential with and without core electrons may function as a slight amount of symmetry breaking.
Therefore, these calculations will represent a lower limit of surface sensitivity.
In  Fig.~\ref{fig:buriedlayer}(a), the total SHG signal as a function of energy can be seen to decrease in accordance with expectations.
Using a definition of layers as two layers per unit cell, it is not surprising that layers 5 \& 6 have a closer signal to layers 7 \& 8 across all energies than layers 1 \& 2 due to the absence of surface effects.

The more surface specific informative analysis comes from comparing the various layers' SHG signal to that of the most buried layers as a function of energy, Fig.~\ref{fig:buriedlayer}(b). 
Here we find that the topmost layers provide a SHG response that is over an order of magnitude larger than the more buried layers near the carbon $K$-edge at 300\,eV, thus indicating a distinct surface dominance of the signal near resonance.
While this trend reduces by 350\,eV, the overall SHG signal is still very surface dominant and emanates from the top 4 most layers of the slab.
However, by 1000\,eV and particularly by 3000\,eV (where each set of layers' ratio approaches unity) the effect of the surface as compared to the bulk is lessened. 
At 3000 eV, the difference in $\chi^{(2)}$ between layers 1 \& 2, and 3 \& 4 is within error, so the increase in relative intensity of 3 \& 4 cannot be assigned as meaningful.
While, to some degree the definition of 'surface sensitive' will always be arbitrary, any reasonable definition would consider 3000\,eV and likely 1000\,eV as functionally a bulk probe. 
While the specific energies at which bulk dominance of the SHG signal occurs will be entirely material dependent and have to be verified on a case by case basis, we anticipate in general that it will occur around or just before the onset of tender x-ray ($\sim$1 keV) regime in most materials in the absence of other core resonances as most materials will have d-spacings comparable to or larger than that of diamond.
Likewise, near any resonances or half-resonances in the soft x-ray regime (<\,1 keV), we expect in general the SHG signal to have an increase in surface sensitivity, but again specific materials/edges will have to be investigated individually.

%\section{Conclusions}

The surface sensitivity of diamond in the x-ray regime was investigated both analytically and computationally within VG-RT-TDDFT.
Analytically for diamond, it was shown that one would expect SHG to be bulk dominant above 3500\,eV and potentially at significantly lower energy.
To further investigate these results, computational calculations which include the full multipole expansion were undertaken as a function of energy in three different ways.
Irrespective of the method applied, the data remains consistent, showing a transition to largely bulk sensitive probe by 1000\,eV, and to largely bulk dominated by 3000\,eV.
However, right near the absorption edge of diamond, SHG remains surface dominant.
While the x-ray SHG response becomes increasingly bulk-dominated on the scale of a keV, the analytical trends and VG-RT-TDDFT results are consistent that the probe is surface dominant near the C $K$-edge indicating that x-ray SHG is powerful tool for investigating surfaces and buried interfaces so long as the probe is soft x-ray and close to a resonant edge.

This work supported by the U.S. Department of Energy, Office of Science, Basic Energy Sciences, Chemical Sciences, Geosciences, and Biosciences Division under Contract DE-SC0023397 and DE-SC0023355. This research used resources of the National Energy Research Scientific Computing Center, a DOE Office of Science User Facility supported by the Office of Science of the US DOE, under contract no. DE-AC02-05CH11231, using NERSC award BES-ERCAP0031431 and BES-ERCAP0035240.
This work also used the Expanse supercomputer at the San Diego Supercomputing Center through allocation  PHY210131 from the Advanced Cyberinfrastructure Coordination Ecosystem: Services \& Support (ACCESS) program, which is supported by NSF Grant No. 2138259, No. 2138286, No. 2138307, No. 2137603, and No. 2138296.

%\section{References}
\bibliography{refs}
\end{document}